\documentclass{article}
\usepackage{authblk}
\usepackage{amsmath,amssymb}
\usepackage{graphicx}
\usepackage{color}
\usepackage{subfigure}
\usepackage{url}
\usepackage{amsfonts}
\usepackage{longtable}
\usepackage{isomath}
\usepackage{lipsum}
\usepackage{widetext}
\usepackage{pdfpages}
\makeatletter
\def\figurename{Figure}
\def\thefigure{\@arabic\c@figure}
\def\fnum@figure{\figurename \ S\thefigure}
\makeatother

\begin{document}


# A 16-bit Coherent Ising Machine for One-Dimensional Ring and Cubic Graph Problems

Kenta Takata[1,2,3,*], Alireza Marandi[1,2,4], Ryan Hamerly[4], Yoshitaka Haribara[1,2,3], Daiki Maruo[1,2,3], Shuhei Tamate[1,2], Hiromasa Sakaguchi[1,2,3], Shoko Utsunomiya[1,2] and Yoshihisa Yamamoto[1,2,4,†]

[1] *ImPACT, Japan Science and Technology Agency, Gobancho 7, Chiyoda-ku, Tokyo 102-0076, Japan*

[2] *National Institute of Informatics, Hitotsubashi 2-1-2, Chiyoda-ku, Tokyo 101-8403, Japan*

[3] *Department of Information and Communication Engineering, The University of Tokyo, Hongo 7-3-1, Bunkyo-ku, Tokyo 113-8654, Japan*

[4] *E. L. Ginzton Laboratory, Stanford University, Stanford, California 94305, USA*

[*]Current email address: takata.kenta@lab.ntt.co.jp
[†]yyamamoto@stanford.edu

**Many tasks in our modern life, such as planning an efficient travel, image processing and optimizing integrated circuit design, are modeled as complex combinatorial optimization problems with binary variables. Such problems can be mapped to finding a ground state of the Ising Hamiltonian, thus various physical systems have been studied to emulate and solve this Ising problem. Recently, networks of mutually injected optical oscillators, called coherent Ising machines, have been developed as promising solvers for the problem, benefiting from programmability, scalability and room temperature operation. Here, we report a 16-bit coherent Ising machine based on a network of time-division-multiplexed femtosecond degenerate optical parametric oscillators. The system experimentally gives more than 99.6 %**

1

of success rates for one-dimensional Ising ring and nondeterministic polynomial-time (NP) hard instances. The experimental and numerical results indicate that gradual pumping of the network combined with multiple spectral and temporal modes of the femtosecond pulses can improve the computational performance of the Ising machine, offering a new path for tackling larger and more complex instances.

2

Many combinatorial optimization problems[1] belong to the complexity classes, called NP-complete and NP-hard[2], and it is believed that they require a computation time scaling exponentially or faster with the number of input variables (problem size). Each problem in the NP-complete class has reducibility from the others within a polynomial time, and hence the classification has broad applications. Thus, an efficient scheme to compute these problems has been extensively searched for.

One of the most popular problems in such classes is to search for a ground state of the Ising Hamiltonian[3]

$$H = -\sum_{i<j}^{N} J_{ij}\sigma_i\sigma_j \ . \qquad (1)$$

Here, $\sigma_i = +1, -1$ is a normalized eigenvalue of a spin 1/2 particle and called Ising spin, and $J_{ij}$ is the sign and magnitude of magnetic interactions. This problem is important to understand mysterious properties of spin glasses and magnetic disorders. To this end, some quantum systems for simulating the Ising model have been developed[4-8]. However, the task of finding a ground state of three-dimensional Ising spin systems has been proven to be NP-hard, while its decision problem version is NP-complete[9,10]. Other NP-complete problems, such as the MAX-CUT problem, graph partitioning problem and some families of SAT problems can be reduced to the Ising problem[11-13].

Meta-heuristic algorithms have been vastly studied to attack this intractable problem. Simulated annealing (SA) [14] is one of the most prevalent and successful methods in practice. Quantum annealing (QA) [15,16] has been proposed as a method which can potentially give better solutions than SA. The hardware to implement QA has also been recently developed[17-19], and its true performance is under consideration[20]. Adiabatic quantum computation[21,22] is closely related to QA and based on the adiabatic evolution of the Hamiltonians, for which some attempts[23,24] are made to avoid the notorious transition[13,25] from an initially prepared ground state to an excited state mainly due to the closing energy gap between them.

We have recently proposed novel computational systems for the Ising problem[26-31], called coherent Ising machines, based on networks of optical oscillators with mutual injections. In the system, an Ising spin is represented by an optical degree of freedom in each oscillator mode such as polarization or phase. Mutual injections between the oscillators induce in-phase and out-of-phase interference which emulates the spin-spin interactions. Here, the Ising Hamiltonian is mapped onto the effective loss of the mutually coupled oscillator network. When a pumping energy is gradually increased to the system from below to above the oscillation threshold[28], the gain of each oscillation mode rises gradually. Accordingly, only one mode with the minimum loss can oscillate first and the other modes are suppressed. Thermal noise is safely neglected in optical systems, thus a low-temperature artificial spin system will be realized in the machine. As SA and QA, coherent Ising machines are considered as ones of the approximation algorithms with finite and polynomially scaling resources, which do not always guarantee the exact solution. Their computational potentials, including the possibility of their performance improvement by quantum effects, have been being theoretically investigated[32, 33].

Degenerate optical parametric oscillators (DOPOs) with the identical pump phase reference[30] are suitable for realization of binary artificial spins. A DOPO shows non-equilibrium phase transition in the distribution function for the anti-squeezed quadrature amplitude $\hat{x} \equiv (\hat{a}+\hat{a}^{\dagger})/2$. In a DOPO pumped below the oscillation threshold (Fig. 1 (a)), the fluctuation in $\hat{x}$ is magnified, while that for the orthogonal component (squeezed amplitude $\hat{p} \equiv (\hat{a}-\hat{a}^{\dagger})/2i$) is suppressed. When the pumping power gets above the threshold (Fig. 1 (b)), the distribution $P(x)$ is divided into the two parts that are completely out-of-phase. Here, the state synchronizing the pump is written as $\{|\theta=0\rangle\} \equiv \{0\}$, and that with the opposite phase denoted by $\{|\theta=\pi\rangle\} \equiv \{\pi\}$. The Ising machine assigns $\sigma_i = +1$ to $\{0\}_i$ and $\sigma_i = -1$ to $\{\pi\}_i$ in the $i$-th DOPO. A time-division multiplexed

DOPO network contains a temporal sequence of signal pulses in a single ring resonator and the mutual coupling is introduced by their interference (Fig. 1 (c)). The number of optical pulses hence can scale with the length of ring cavity and the repetition rate of the pump pulses. An experimental demonstration[31] of a 4-pulse system has shown error-free computation on a frustrated 4-spin ($N = 4$ MAX-CUT-3) problem in 1000 trials. Here, a numerical simulation based on step-function (abrupt) pumping did not reproduce the performance which the experiment showed. It was indicated that gradually introduced pumping could explain the experiment, but further theoretical and experimental evidences with larger and more complex problems were desired for the confirmation. Another difference to the previous simulation is that the implemented system is based on ultrashort pulses with a massive number of frequency modes. A theoretical model and experimental data have been needed to investigate possible multi-mode effects on the machine performance.

In this letter, we show another experimental demonstration of a coherent Ising machine using sixteen telecom-band DOPO pulses resonating in a single ring cavity and pulse-to-pulse phase-controlled interference via three optical delay lines. We have constructed a new system including a pump laser with a higher pulse repetition rate, more stable cavity locking equipment and measurement apparatus with broader bandwidths. An instance of NP-hard MAX-CUT problem on a cubic graph (3-regular graph) is realized, and it has many metastable suboptimal solutions (local minima) which are not included in the previous experiment. In addition, one-dimensional ferromagnetic and anti-ferromagnetic Ising ring problems are programmed for benchmarking. A slow pumping schedule is successfully implemented, and the dynamics of the signal pulses is measured in a long time scale with a high-speed detector. Computation with negligible error rates for 1000 or more successive trials is achieved. We also compare the experiment with numerical simulations of abruptly pumped single-mode DOPOs, gradually pumped single-mode DOPOs and abruptly pumped multi-mode DOPOs based on temporal Hermite-Gauss pulses. It shows that the

gradual pumping and multi-mode dynamics support the high performance in the experiment. Fine signal dynamics in the experimental data suggests that enhanced state-to-state tunneling due to the multi-mode operation protects the system from fluctuation coming from experimental defectiveness.

**Result**

   **Experimental setup.** The experimental system is based on phase-locked femtosecond DOPOs[34, 35] running in a single ring cavity (Fig. 2 (a)). The pump laser (Laser Quantum, Taccor 10s) is a mode-locked titanium-sapphire laser emitting pulses with a temporal duration of 14 fs, a central wavelength of 794 nm and a repetition frequency of 1 GHz. A mechanical chopper (CP) is placed in the cavity and periodically blocks the pump beam, enabling independent sessions of computation. The pump laser beam has a spot size of about 1 mm at its output port, thus the chopper gives a finite rise time for the average pump power, depending on the shape of the blade and the switching frequency. It achieves slow increase in the signal quality factor and the pumping power. Both of them gradually raise the net gain and help the system to reach a lowest-loss state. The round-trip time of the ring cavity composed of four mirrors (M1 - M4) is 16 ns, and the pump repetition period is 1 ns, resulting in sixteen independent signal pulses per round trip. The dielectric mirror M1 transmits more than 90 % of the pump power while reflects 99.8 % of the signal. The mirrors M3 and M4 are gold-coated concave mirrors with a radius of curvature of 50 mm. These two mirrors focus the pump and signal beams in the nonlinear crystal and achieve a large parametric gain. The nonlinear crystal is a 1-mm long MgO-doped periodically poled lithium niobate (PPLN) with a poling period of 20.4 μm and put at the Brewster angle, providing degenerate type 0 (e + e → e) quasi phase matching (parametric gain). All the input and output couplers (ICs and OCs) are 1 mm-thick plate beamsplitters with AR coating on one side and 10% reflective coating on the other side. PD1 is a slow detector for monitoring the average signal power and is also used for cavity stabilization. The main servo controlling system (Controller 1, TEM Messtechnik, LaseLock Digital)

places a PID feedback, using a top-of-fringe locking scheme[36], stabilizing the cavity length via the piezoelectric transducer (PZT). The mechanical chopper (CP) enables automated and independent computation runs. However, the resulting abrupt changes in the PD output induce some instability in the cavity locking by Controller 1. We thus use another servo controlling circuit with a very low bandwidth (Controller 2). Here, we put another small and relatively fast modulation of 20 kHz on the fast PZT channel of the pump laser, and the circuit gives the proportional feedback with the fast error signal to the slow PZT channel. The whole system is surrounded by a cardboard enclosure to avoid the fluctuation due to airflow and sound. See also Methods and Supplementary Information.

Couplings between the DOPO pulses are implemented by the three pairs of OCs and ICs in the cavity (Fig. 2 (b)). Each of them has a reflection of $R \sim 10\%$ in power for DOPO pulses. It means that each pair of OCs and ICs provides ~ 10% of nominal field couplings between the pulses, while the additional loss in the system degrades the ratio. The effective coupling estimated with the change in the oscillation threshold by opening delay 1 is 7.5 %. Delay 1 lags the pulses by 1 ns and gives the couplings from the *n*-th to (*n*+1)-th OPO, i.e. $J_{n\,n+1}$. Likewise, delay 2 (15 ns) and delay 3 (8 ns) provide the couplings $J_{n+1\,n}$ and $J_{n\,n+8} = J_{n+8\,n}$, respectively. The sign of the coupling $J_{ij}$ is controlled by the PZT stage that tunes the optical path length of each delay line. The oscillation threshold of the system with two output couplers for PD1 and the readout interferometer is 113 mW, and the threshold including the three delay lines is 334 mW. The unequal-arm interferometer is used for the measurement of the relative phases between adjacent signal pulses via both slow and fast detectors. High-level and low-level pulses at the fast detector suggest $\sigma_{i+1} = +\sigma_i$ and $\sigma_{i+1} = -\sigma_i$, respectively, thus the output pulse pattern allows us to measure the artificial spin configuration. The servo controllers of the same model as Controller 1 also lock the interferometer and delay lines using the interference of the pump pulses, although they are omitted from the figure for simplicity. Here,

blocking of the pump beam and hence the electrical feedback by the chopper interrupts the stable couplings. We thus use a narrow chopping blade to reduce the off-pumping time as much as possible. As the previous work[31, 37], we have obtained the data showing that each pulse randomly takes either of binary phases, 0 or π, when the pulses are decoupled. Also, we have assured that introducing a single delay line (delay 1) and modulating its coupling phase can switch between the same-phase state $\sigma_i \sigma_{i+1} = +1$ and alternating-phase state $\sigma_i \sigma_{i+1} = -1$ (See Supplementary information).

**Performance of the system.** Table 1 summarizes the performance of the 16-pulse coherent Ising machine for ferromagnetic and antiferromagnetic Ising rings, and a cubic graph where each node is connected with three other ones. The two one-dimensional ring instances involve only delay 1 and 2, with delay 3 blocked. Opening all the three delay lines gives the couplings aligned in the cubic graph called Möbius ladder. The anti-ferromagnetic couplings on such cubic graphs represent NP-hard instances, equivalent to the MAX-CUT-3 problem. Fig. 3 depicts the fast-detector output signals corresponding to one of the ground states of the target instances illustrated below the graphs. Both instances of ferromagnetic and anti-ferromagnetic rings have only two degenerate ground states, which are pairs of totally opposite phase state configurations with the same-phase (Fig. 3 (a)) and alternating-phase orders (Fig. 3 (b)), respectively. While these instances do not have any local minima, they include a series of first excited states with the two-domain structures, where moving the domain wall does not change the cost function. The sixteen ground states of the MAX-CUT-3 problem in the cubic graph with homogeneous anti-ferromagnetic couplings have two domains of eight anti-phased pulses and two frustrated parts at the domain boundaries, corresponding to the high-level pulses in Fig. 3 (c). 34 local minima correspond to the states strongly reflecting the couplings of either the ring or the diagonals. The eight pulse patterns that are cyclic permutations of Fig. 3 (c) also correspond to the ground states, although we do not distinguish them because of the absence of a time reference in our measurements.

We continuously measure and record the output pulse pattern for each instance to estimate the probability of finding a ground state. We use a single narrow blade for the chopper with a frequency of 20 Hz. The rise time (10 % - 90 % of the maximum) of the pump power is 206 μs corresponding to ~12,800 round trips for the 4.8 m ring cavity. For the ferromagnetic and anti-ferromagnetic 1D ring, we have got 100 % and 99.6 % success rates out of 1,000 runs. For the anti-ferromagnetic cubic graph, the success rate was 100 % in 2,000 trials. The small error rate for the anti-ferromagnetic 1D ring case could be due to the experimental incompleteness such as the beam alignment and mechanical stability of the system.

To examine the impact of the gradual pumping and multimode effect, we simulated the system using three theoretical models. The first model is based on the abruptly pumped single-mode DOPO network with discrete gain and coupling processes[38, 39]. It performs stepwise introduction of a constant pumping power of 2.7 times the oscillation threshold ($I_{th}$), which is comparable with the experiment, to the system in the vacuum state. The second model incorporates a linearly scheduled pumping to the first model. The pump rate is increased from $I_{th}$ to 2.7 $I_{th}$ with 10000 round trips, so that its slope is comparable with that in the experiment. As in the laser case[28], it is expected that a slower schedule leads to a higher success probability at the expense of a longer computation time. The third model (described in Supplementary Information) deals with multimode DOPOs[40] with an abrupt pumping of 1.1 $I_{th}$. Here, five temporal Hermite-Gauss modes are assumed for each pulse. The coupling ratios are all the same as the experimental value (7.5 %). The simulation result is shown in Table 1. Although all the models solve the cubic graph problem, the first model with 10000 round trips gives success rates of only about 60 % in the 1-D Ising ring problems. We have assured that 300 round trips make these rates converge, thus their further improvement with more round trips cannot be expected. It means that the system of abruptly pumped single-mode DOPOs is likely to be trapped in local minima and not consistent with the experimental result. Meanwhile, the success

9

probabilities by the second model converge at unity, 2000 round trips after crossing the oscillation threshold of the individual DOPO. Also, the multimode tunneling accelerates the search for a ground state. Only 200 times the cavity lifetime, corresponding to about 240 round trips are enough for the third model to achieve error-free computation of the problems considered. The experimental performance supports the enhancement of the success rates due to these effects. In terms of the stability of the current locking scheme, it is difficult to change the pumping schedule and hence identify which effect is dominant. Finer controllability of the system is a challenge for the future.

**Long-term pulse measurement.** The detailed dynamical behavior of the OPO network is studied. Measured data indicates how the tunneling and gradual pumping contribute to the computing process of the machine. Fig. 4 presents an example of the signal mode interferometer outputs in the presence of a chopper, for the cubic graph problem. A metallic chopping fan with multiple blades is used, and the pump rise up time is 672 μs. The gradual turnover is firstly demonstrated by the signal envelope which increases linearly in time from $t \approx 70$ μs, corresponding to the effective oscillation threshold of the DOPO network (Fig. 4 (a)). The oscillatory behaviour is seen until $t \approx 115$ μs, and it gradually slows down towards the stable final state. A zoomed plot (Fig. 4 (b)) shows that the system finds a ground state at an early peak in the oscillation. However, the state significantly changes at a certain point, and the envelope of the output power decreases afterwards (Fig. 4 (c)). It means that the system lets go of the ground state by the fluctuation which would come from the excess noise of the pump laser. Although, the excited state which the system reaches has a larger loss, and the gain is limited by the gradual pumping. As a result, the system can damp the signal pulses and oscillate a ground state again (Fig. 4 (d)). Every peak of the output for $t > 80$ μs, with an interval between 100 and 200 round trips, includes the pattern for ground states. This is not consistent with the slow convergence of the success probability by the simulation of the gradually pumped single-mode DOPO network, and suggests the accelerated computation by the multimode effect. As the signal

10

gain increases in time, the relative magnitude of the fluctuation falls. This leads to the slowing of the oscillation in the output envelope. When the pulses have enough amplitude, the switching between different configurations gets impossible and the system stably holds the ground state (Fig. 4 (e)).

## Discussion

Here, we discuss the experimental defectiveness. Errors mostly come from the misalignment of the beams for the couplings. Here, we have to correctly overlap the signal beams not only spatially but also temporally. Pulses have fine structures of optical cycles, thus the interfered pump power output from the IC of each delay line can have many extremal points, at which the electric feedback locks the system. However, the duration of the pump pulses in our system corresponds to just 7.6 cycles, thus it is relatively easy to obtain the best condition by searching for the maximum average cavity signal power, inversely proportional to the photonic loss.

We also refer to the erroneous states which appear when the system is not optimally tuned, in terms of the temporal locking point (see also Supplementary Information). In the 1-D ring problems, they are prone to be a series of first excited states containing two magnetic domains. In the cubic graph case slightly off the best condition, we have observed two local minima out of 1000 runs. The effect of the experimental incompleteness tends to be along with general errors in approximate optimization algorithms. In larger systems, it will be more important to distinguish the fundamental errors of the machine from the imperfection of the experimental condition.

## Conclusion

In conclusion, we have presented an experimental demonstration of a 16-pulse coherent Ising machine. The precisely controlled system avoids excited states and finds ground states of one-dimensional ring and cubic graph instances with sufficiently small error rates, supporting the advantage of the gradual pumping and multi-mode operation. The dynamics of the interference signal reveals multiple destruction and construction of ground states under the experimental

fluctuation and limited gain. Multimode tunneling is indicated in terms of the fast search for ground states. Combination of building a ring oscillator in a fiber cavity and implementing effective pulse-to-pulse couplings with a measurement-feedback technique[38] is expected as a practical way to improve the programmability and scalability of the machine. Another important direction is to examine if possible quantum effects[32] in the system improve the probability to find a ground state. Our simulation work here is based on semiclassical models, thus estimation of the quantum speedup is beyond the scope of this paper. However, very recent theoretical work[33] indicates that our 16-bit Ising machine might contain quantum correlation which restricts undesired excited states, even under some loss in the cavity. Experimental confirmation of such quantum effects is left for future study.

## Methods

**Operation of optical parametric oscillator.** The feedback control of the cavity length enables the continuous operation. The cavity length corresponding to the degenerate operation results in the highest slope efficiency and a single-peak output spectrum. The signal pulses at degeneracy have a central wavelength of 1574 nm, a 3-dB spectral width of 80 nm, pulse duration of 80 fs and a pulse repetition rate of 1.0096 GHz. See also Supplementary Information.

**Numerical simulation.** In the numerical simulation, we adopt the single-mode and multi-mode theory and assign a c-number field variable for each mode. The single-mode model is based on Ref. 40. Numerical integration is based on the fourth-order Runge-Kutta method. Discrete steps such as parametric amplification, out-coupling and input-coupling including vacuum fluctuation are sequentially applied according to the experimental setup. The phase states of the fields are read out after 10,000 round trips for a single run. 1,000 trials are used for estimation of the success probability. The multi-mode theory exploits five Hermite-Gauss modes for a single pulse. 200 times

the cavity lifetime (about 240 round trips) were enough to find a ground state for 1,000 runs. Details are presented in Supplementary Information.

**MAX-CUT problem.** MAX-CUT problem is a kind of bipartition problems for a graph $G = \{V, E\}$ composed of a set of vertices $V$ and that of edges $E$ connecting two vertices. A partition of $V$ into two disjoint subsets $V_1$ and $V_2$ is called a *cut*. A *cut size* is the number of edges between vertices of different subsets of a cut ($V_1$ and $V_2$). Here, MAX-CUT problem is defined as finding the maximum cut size for a given graph and belongs to the NP-hard class.

## Acknowledgements

## Author contributions

K. T., A. M., S. U. and Y. Y. organized the project. K. T. and A. M. designed and carried out the experiment. R. H. developed the theoretical model. R. H. and Y. H. performed numerical simulation. K. T. wrote the manuscript. Y. H. supported the supplementary materials. D. M., S. T., H. S. and S. U.

supported the experiment. S. T. supported the theoretical analysis. All of them commented on the manuscript.

## Competing financial interests
The authors declare no competing financial interests.

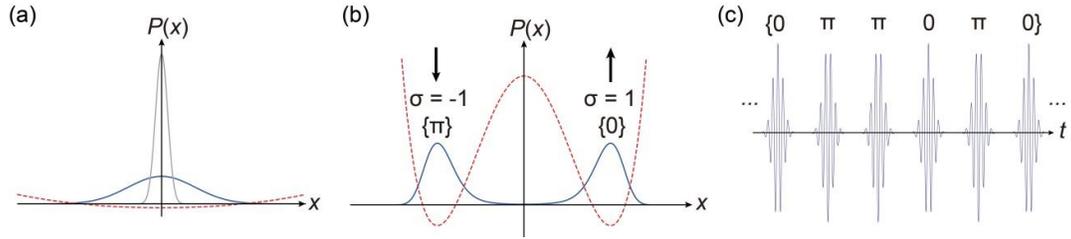

**Figure 1 | Optical binary bits based on the anti-squeezed quadrature amplitude *x* in degenerate optical parametric oscillators.** Distribution function (solid blue line) and effective potential (red dashed line) for the quadrate amplitude *x* (a) below and (b) above the oscillation threshold. Gray line in (a) is that for the vacuum state. Oscillating fields with discrete phases 0 and $\pi$ map the spin variables $\sigma = +1$ and $-1$. (c) Temporal sequence of the optical bits in a time-division multiplexing system. The repetition rate and the length of the ring cavity for the pulses determine the number of spins available for computation.

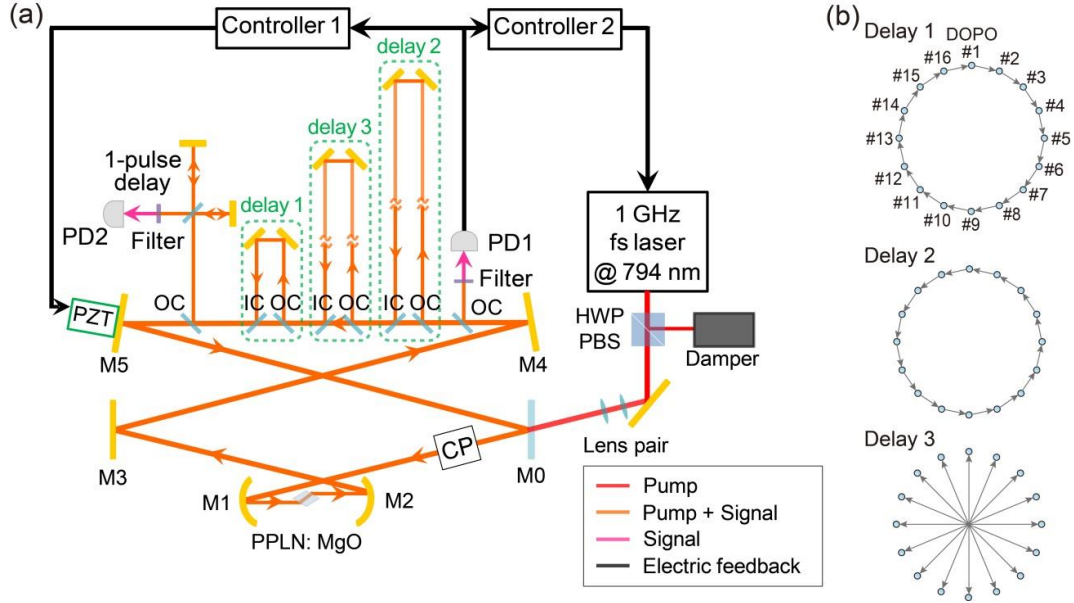

**Figure 2 | Experimental setup.** (a) The pump laser is a Ti:sapphire femtosecond pulse laser with a central wavelength of 794 nm. The OPO ring cavity is 4.8 m long, and the round trip time for the pulses is 16 ns. Three delay lines implement the optical coupling between the DOPO pulses. An unequal-arm Michelson interferometer measures the relative phases of adjacent signal pulses. Two servo controllers are used to stabilize the OPO under the operation of the mechanical chopper. (b) Couplings introduced by each optical delay line. Delay 1 (1 ns) couples adjacent pulses in the forward direction, while that of Delay 2 (15 ns) is backward, feeding the pulses back to those of the next round trip. Delay 3 (8 ns) introduces the mutual couplings between the pulses which are half round trip (8 pulses) away. HWP: half wave plate, PBS: polarizing beamsplitter, CP: chopper, M: mirror, IC: input coupler, OC: output coupler, PD: photodetector, PZT: piezoelectric transducer.

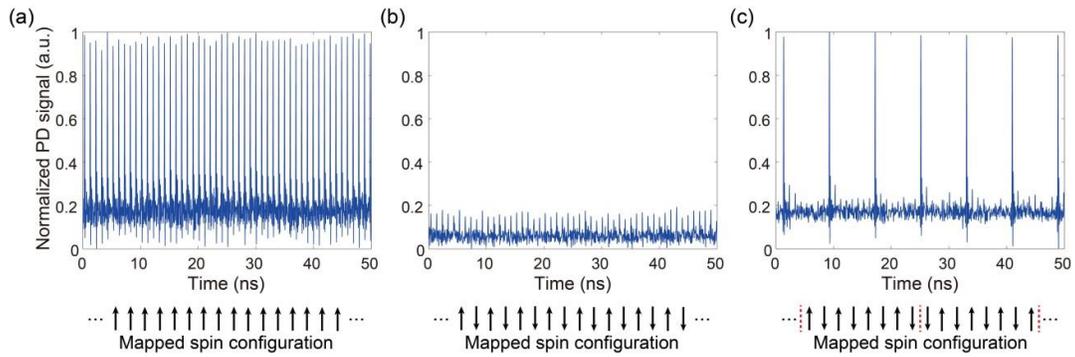

**Figure 3 | Interferometer output signals corresponding to the ground states.** (a) Complete same-phase order for the one-dimensional ferromagnetic ring instance. (b) Complete alternating-phase order for the one-dimensional anti-ferromagnetic ring instance. (c) Answer to MAX-CUT; Two alternating-phase (anti-ferromagnetic) domains containing eight pulses, with frustration at the boundaries for a cubic-graph instance. Corresponding Ising spin configuration is shown below each graph. Red dashed lines in (c) display the domain boundaries.

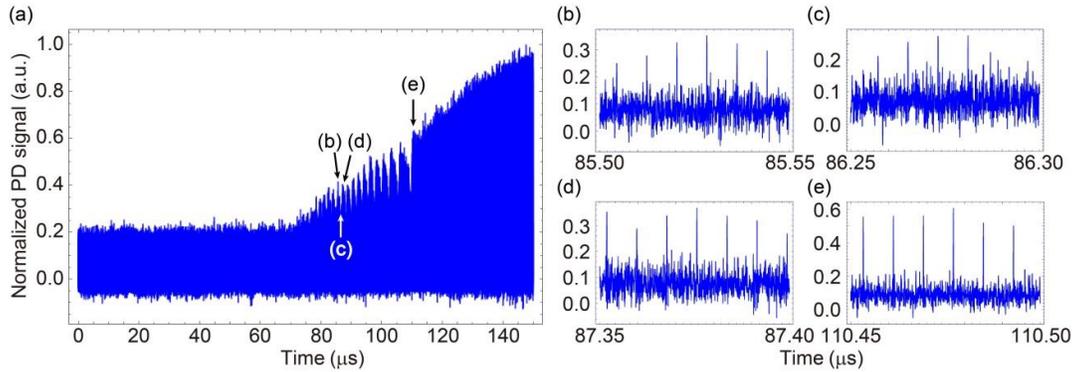

**Figure 4 | Dynamics of the gradually pumped coherent Ising machine in terms of the interferometer output around the oscillation threshold.** The average pump power is 2.7 times of the oscillation threshold and the pump riseup time is 672 μs. (a) A nearly linear increase of the signal envelope shows the realization of gradual pumping with a mechanical chopper. The oscillation starting from $t \sim 80$ μs is due to the destruction and construction of ground states under the pump fluctuation and gradual pumping. (b) A ground state of the cubic graph problem is reached at an early peak ($t \sim 85.5$ μs). (c) However, it suddenly jumps to another excited state and decays at once around $t = 86.3$ μs. (d) The pulse configuration for ground states can be formed at the next peak ($t = 87.3$ μs). (e) After the slowing cycles of oscillation along with the increase of the pump power, a stable ground state is formed at $t \sim 110$ μs.

**Table 1 | Summary of the performance of the 4-bit and 16-bit coherent Ising machine.**

|  | One-dimensional ring, ferromagnetic | One-dimensional ring, anti-ferromagnetic | Cubic graph, anti-ferromagnetic ($N = 16$) |
|---|---|---|---|
| Coupling | $J_{i,i+1} = J_{i+1,i} = 1$ (delay 1, 2) | $J_{i,i+1} = J_{i+1,i} = -1$ (delay 1, 2) | $J_{i,i+1} = J_{i+1,i} = -1$ $J_{i,i+8} = J_{i+8,i} = -1$ (delay 1, 2, 3) |
| Number of ground states | 2 | 2 | 16 |
| Example of ground states $\{\sigma_i\}$ | {1, 1, 1, 1, 1, 1, 1, 1, 1, 1, 1, 1, 1, 1, 1, 1} | {1, -1, 1, -1, 1, -1, 1, -1, 1, -1, 1, -1, 1, -1, 1, -1} | {1, -1, 1, -1, 1, -1, 1, -1, -1, 1, -1, 1, -1, 1, -1, 1} |
| Number of local minima | 0 | 0 | 34 |
| Ground state energy | -16 | -16 | -20 |
| Average pumping power | 1.00 W | 900 mW | 1.00 W |
| Experimental performance | 1000/1000 | 996/1000 | 2000/2000 |
| Simulation 1 (single-mode, 10000 round trips) | 604/1000 | 588/1000 | 1000/1000 |
| Simulation 2 (gradual pumping, 10000 round trips) | 1000/1000 | 1000/1000 | 1000/1000 |
| Simulation 3 (multimode, 200 round trips) | 1000/1000 | 1000/1000 | 1000/1000 |



\title{Supplementary Information for A 16-bit Coherent Ising Machine for One-Dimensional Ring and Cubic Graph Problems}

\author{Kenta Takata$^{1, 2, 3,}$\thanks{takata.kenta@lab.ntt.co.jp} , Alireza Marandi$^{1, 2, 4}$, Ryan Hamerly$^{4,}$, Yoshitaka Haribara$^{1, 2, 3}$, Daiki Maruo$^{1, 2, 3}$, Shuhei Tamate$^{1, 2}$, Hiromasa Sakaguchi$^{1, 2, 3}$, Shoko Utsunomiya$^{1, 2}$ and Yoshihisa Yamamoto$^{1, 2, 4}$}
\affil{$^{1}$ImPACT, Japan Science and Technology Agency, \\ Gobancho 7, Chiyoda-ku, Tokyo 102-0076, Japan}
\affil{$^{2}$National Institute of Informatics, \\ Hitotsubashi 2-1-2, Chiyoda-ku, Tokyo 101-8403, Japan}
\affil{$^{3}$Department of Information and Communication Engineering, \\ The University of Tokyo, Hongo 7-3-1, Bunkyo-ku, Tokyo 113-8654, Japan}
\affil{$^{4}$E. L. Ginzton Laboratory, Stanford University, \\ Stanford, California 94305, USA}

\maketitle

\section{\label{multimode}Multimode tunneling effects in a DOPO Ising machine}
\subsection{Equations for signal fields and Hermite function expansion}
The time-division multiplexed degenerate optical parametric oscillators (DOPOs) sometimes exhibit higher order pulse formation due to a large number of longitudinal cavity modes. We review the formulation according to the Ref. \cite{paper:PTFV10}. 
The signal field operator at the position $z \in [0, L)$ inside the cavity of its length $L$ can be written as
\begin{equation}
\hat{E_\mathrm{s}}(z, t) = \sum_{m}i \mathcal{E} \hat{s}_m(t) e^{i\omega_{\mathrm{s},m}(z/\nu - t)} + \mbox{H.c.},
\end{equation}
where $\mathcal{E}$ is the single photon field amplitude and $\hat{s}_m(t)$ is the signal annihilation operator. The signal frequency of $m$-th longitudinal mode $\omega_{\mathrm{s},m} = \omega_0 + m\Omega\ (m\in \mathbb{Z})$ has an interval of the cavity free spectral range $\Omega$ centered at $\omega_0$. 
After the pump adiabatic elimination, the signal annihilation operator obeys the following equation:
\begin{eqnarray}
\frac{d\hat{s}_m}{dt} & = & -\gamma_\mathrm{s} \hat{s}_m - i(\Delta + m\Delta \Omega) + \sqrt{2\gamma_\mathrm{s}} \hat{s}_{{\rm in},m} \nonumber\\&&
+ K \sum_q f_{m,q}\hat{s}_q^\dagger \hat{p}_{{\rm in},m+q} - \frac{K^2}{4} \sum_{q, n}{ f_{m,q} f_{n,m+q-n} \hat{s}_q^\dagger\hat{s}_n \hat{s}_{m+q-n}}.
\end{eqnarray}
Here, the signal photon decay rate $\gamma_\mathrm{s}$ is equal to $\Omega T_\mathrm{s} / 4 \pi$ with the transmittance (cavity loss) $T_\mathrm{s} \ll 1$, the second term of the right hand side is the detuning $\Delta$ and pump timing mismatch $m\Delta \Omega$, and the coupling constant $\kappa$ is included in $K = 2 \kappa \sqrt{L/c}$. 
The phase-mismatching factor $f_{m, q}$ between two signal modes $m, q \in \mathbb{Z}$ is a sinc function of the phase-mismatch angle.

The parametric coupling between two signal modes is given by a matrix $\mathcal{L}_{m, q} = f_{m, q} \alpha_{m+q}$ composed of the phase-mismatching factor $f_{m, q}$ and the complex spectral component $\alpha_m$. 
At the continuous limit, the component of its eigen vectors ${\psi_k}$ are written by Hermite function 
\begin{align}\label{eq:Hermite}
\psi_{k, m} = \frac{1}{\sqrt{k! 2^k \sqrt{\pi} N_\mathrm{s}}} e^{-\frac{1}{2}( \frac{m}{N_\mathrm{s}})^2} H_k\left(\frac{m}{N_\mathrm{s}}\right)
\end{align}
with an Hermite polynomial of order $k$ and the number of signal modes $N_\mathrm{s} = (\Omega \tau_\mathrm{s})^{-1}$.

Now the signal photon annihilation operators can be expanded by 
Hermite functions, 
$
\hat{s}_m = \sum_{k \in \mathbb{N}} \psi_{k, m} \hat{S}_{k}, 
$
and follow equations below by this basis transformation:
\begin{widetext}
\begin{eqnarray}
\frac{d\hat{S}_i}{dt} &=& -(\gamma_s + i\Delta)\hat{S}_i + \sqrt{2\gamma_s}\hat{S}_{\mathrm{in}, i} \nonumber
\\&& \quad - i\Delta\Omega\sum_{j,m}m\psi^*_{i,m}\psi_{j,m}\hat{S}_j + K\sum_{j,m, q}f_{m, q-m}p_{\mathrm{in}, q}\psi^*_{i,m}\psi^*_{j,q - m}\hat{S}^\dagger_j \nonumber\\
&& \quad - \frac{K^2}{4}\sum_{ j, k, l, m, q, n}f_{m, q - m}f_{n, q - n}\psi^*_{i,m}\psi^*_{j,q - m}\psi_{k,n}\psi_{l,q - n}\hat{S}^\dagger_j \hat{S}_k \hat{S}_l \\
\label{eq:signal}
&=&\!: -(\gamma_s + i\Delta)\hat{S}_i +\! \sqrt{2\gamma_s}\hat{S}_{\mathrm{in}, i} \nonumber
\\&& \quad-\! i\Delta\Omega\sum_{j}D_{ij}\hat{S}_j +\! K\sum_{j}G_{ij}(p_{\mathrm{in}})\hat{S}^\dagger_j -\! \frac{K^2}{4}\sum_{j, k, l}L_{ijkl}\hat{S}^\dagger_j \hat{S}_k \hat{S}_l, 
\end{eqnarray}
\end{widetext}
where $D_{ij}$ and $G_{ij}$ denote coefficient for pump timing mismatch and PPLN OPA gain, respectively, while $L_{ijkl}$ denotes signal mode coupling between two coupled modes $(i, j)$ and $(k, l)$. 
In a coherent Ising machine, $N$ DOPOs are mutually coupled each other with given coupling constants \cite{paper:WMWB13}. The set of Eq. (\ref{eq:signal}) including the mutual coupling term is used in the numerical simulation.

\subsection{Dynamics of the multimode tunneling}
\begin{figure}[htbp]
\begin{center}
\includegraphics[width=10cm,trim = 0 0 0 0,clip]{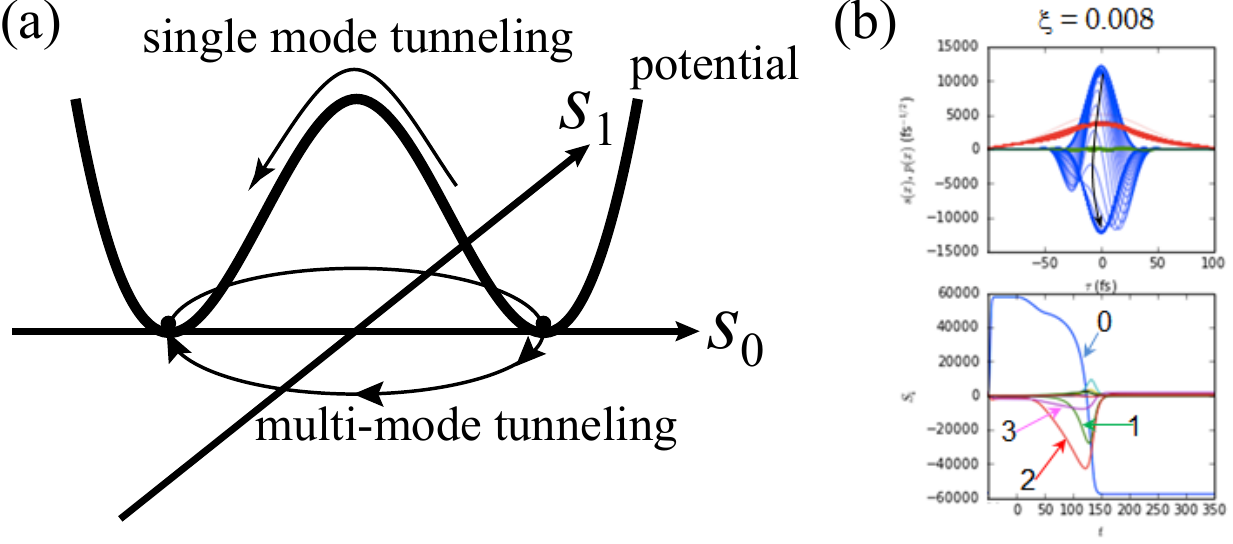} 
\caption{(a)Schematic and (b) Simulated dynamics of the phase flip in a multimode DOPO. Upper panel in (b): temporal profiles of Hermite modes of a DOPO pulse. Lower panel: time evolution of the coefficients of the modes.}\label{fig:MultimodeTunneling}
\end{center}
\end{figure}

Here, we show the picture of the multimode tunneling in DOPOs with Fig. S\ref{fig:MultimodeTunneling} (a).
In the system of single-mode DOPOs, the phase flip of a macroscopic oscillator field requires to pass through the complete cancellation of the mode (the state with zero amplitude), which is unfavorable.
On the other hand, a multimode DOPO can switch between the binary phase states by rotating its phase with the excitation of higher spatio-temporal modes which have different frequencies.
It helps the system escape from unstable states and smoothly find a ground state of the programmed Ising model.
Fig. S\ref{fig:MultimodeTunneling} (b) displays an example of simulated dynamics of the phase flip in a multimode DOPO pulse under mutual injections aligned in the one-dimensional ring, with a uniform amplitude coupling coefficient $\xi = 0.008$.
When the sign of the fundamental mode amplitude $s_0$ is reversed, those for higher modes temporarily rise and keep the intensity of the whole pulse.
Such a transition enhances the tunneling rate of a multimode DOPO compared to a single mode system.
More details will be published elsewhere \cite{prep:Alireza_Ryan}.

\subsection{Simulated performance}
The performance of 16-pulse DOPO Ising machine is numerically evaluated. Here, both the single-mode and multimode treatments of in-phase signals are calculated by fourth-order Runge-Kutta method. Note that the effective number of signal modes in multimode simulation (indexed by $k$ in Eq. (\ref{eq:Hermite})) is assumed to be 5 for each pulse. A stepwise and constant pumping rate of 1.1 $I_{th}$ is introduced at $t = 0$, with $I_{th}$ being the oscillation threshold for the single DOPO.
The phase states of the fields are read out after 200 times the cavity lifetime for a single run, and then 1000 trials are used to estimate the success probability. 
The result, when the coupling constant is $7.5 \%$, is summarized in Table \ref{tab:srate}. 
The Ising problems are performed on three graphs as in the experimental demonstration: the ferromagnetic one-dimensional ring, the anti-ferromagnetic one-dimensional ring, and an anti-ferromagnetic cubic graph. The difference between single-mode and multimode is significant in two (ferromagnetic and anti-ferromagnetic) one-dimensional rings, while the ground states of the cubic graph with $N=16$ is $100 \%$ attainable in both methods. We add the result for the cubic graph with $N=4$, which was experimentally solved in the previous study \cite{paper:MWTB14}.

\begin{table}[h]
\centering
\begin{tabular}{ccccc}
& ferro & anti-ferro & cubic ($N=16$) & cubic ($N = 4$) \\\hline
single-mode & .593 & .599 & 1.000 & 0.930 \\
multimode & 1.000 & 1.000 & 1.000 & 1.000 \\\hline
\end{tabular}
\caption{\label{tab:srate}Success rate in 1000 trials of numerical calculations for Ising problem on three graphs of order $N = 16$.}
\end{table}

\section{Simulation result of the single-mode DOPO network}\label{sec:SingleModeSimulation}
Here, we theoretically show that the gradually introduced pumping can improve the performance of the Ising machine in term of the success probability, by simulating a model based on single-mode DOPOs. For the probabilistic simulation of the Ising machine, we apply a theoretical model where each DOPO sequantially undergoes the input and output, parametric gain and beamsplitter coupling processes. This scheme is basically the same as one described in Ref. \cite{paper:HII16,paper:HYU16}. Here, vacuum fluctuation of the pump and signal modes is introduced at each input and output coupler. Noise terms are not included in the parametric gain process. Also, the squeezed quadrature amplitude $p$ in every DOPO is not explicitly considered in the simulation, because its average value decays in the parametric process.

\begin{figure}[htbp]
\begin{center}
\includegraphics[width=12cm]{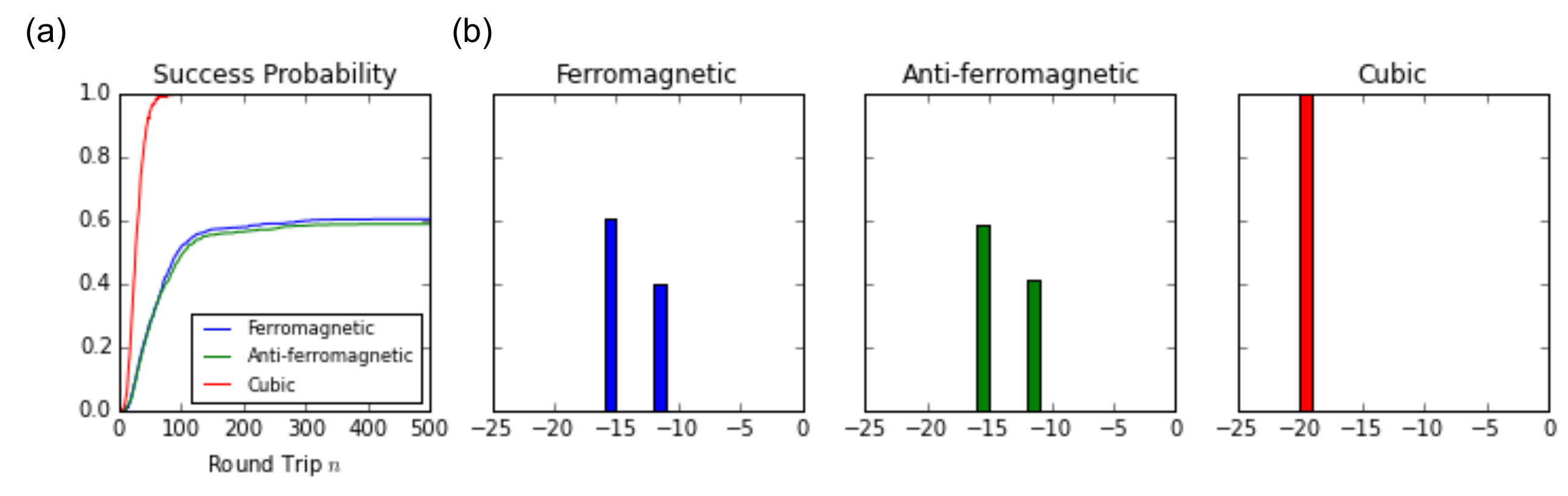}
\end{center}
\caption{(a) The simulated transient success probabilities of the abruptly pumped single-mode DOPO Ising machine for the ferromagnetic ring, anti-ferromagnetic ring and cubic graph problems. (b) The distributions of output states after 10000 round trips, for the problems considered.}\label{fig:SingleModeAbrupt} 
\end{figure}

Fig. S\ref{fig:SingleModeAbrupt} (a) presents the simulated time evolution of the successs rates given by the single-mode Ising machine with a step-function introduction of pumping (abrupt pumping) out of 1000 runs, for the problems used in the experiment. Here, the pump power is 2.7 $I_{th}$. The reflection rate for the mutual injections is 7.5 \%. The success probability for the cubic graph problem converges at unity with about 100 round trips. However, the scores for the one-dimensional ring problems are lower and saturate around 300 round trips. The final values at 10000 round trips are 0.604 and 0.588 for the ferromagnetic and anti-ferromagnetic ring instances. The energy distributions of the final states for these problems are shown in Fig. S\ref{fig:SingleModeAbrupt} (b). The ground energies for the ferromagnetic ring, anti-ferromagnetic ring and cubic graph instances are -16, -16 and -20, respectively. The energy for the erroneous states in the 1-D problems is -12, meaning formation of two domain walls.

\begin{figure}[htbp]
\begin{center}
\includegraphics[width=3.5cm]{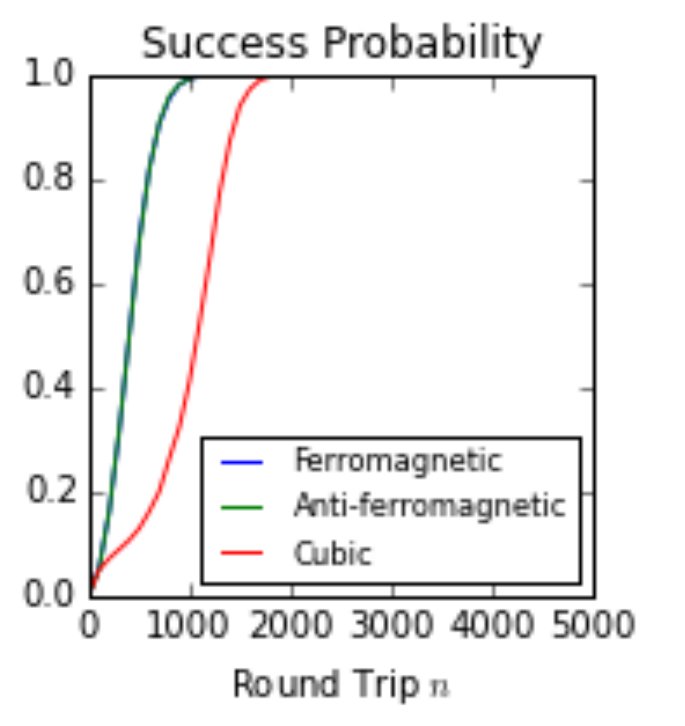}
\end{center}
\caption{Time evolution of the success rates given by the gradually pumped Ising machine. The pump rate is linearly increased from the single DOPO oscillation threshold $I_{th}$ to 2.7 $I_{th}$ with 10000 round trips.}\label{fig:SingleModeGradual} 
\end{figure}

Fig. S\ref{fig:SingleModeGradual} shows the transient of the success probabilities with the Ising machine under the gradual pumping. We apply a linear pumping schedule from $I_{th}$ to 2.7 $I_{th}$ over 10000 round trips. It takes 1000 round trips for the perfomance of 1-D ring problems, and about 1700 round trips for that of the cubic graph problem, to reach 100 \% success. Comparing with Fig. S\ref{fig:SingleModeAbrupt} (a), we can say that the gradual pumping improves the possibility to obtain a ground state of the Ising model of the 1-D ring. Also, the experimental result is compatible with Fig. S\ref{fig:SingleModeGradual}.

We compare the simulation result with the temporal experimental data for the cubic graph problem (Fig. 4), and discuss the accelerated search via the multimode effect. The measured signal in the experiment shows the repeated oscillation and decay of the pulse pattern correspoinding to the ground states, lasting for about 40 $\mu$s. However, such behavior, with repetition rates close to 1 MHz, has not seen in the simulation with both the single-mode and multimode models. Thus, it is attributed to the state disruption by experimental excess noise, which would mainly come from the pump laser. Here, in the case of the single-mode simulation with the slow pumping schedule, it takes about 30 $\mu$s for the success rate to converge at unity (Fig. S2). On the other hand, the patterns for ground states are experimentally observed in output peaks with large magnitude compared with noise, and the intervals of the peaks are a few microseconds at most (Fig. 4(a)). The fast recovery of the solution with a very high probability is not expected only by the gradual pumping, and is considered to be achived with the help of the multimode bit flip.

\section{Operation and properties of the 16-pulse optical parametric oscillator with telecom wavelengths}\label{sec:OPOproperties}
The construction and operation of a pulsed OPO are performed in the two modes. First, we run the system in the \textit{scanning operation}, where the cavity length is scanned over hundreds of micrometers with a periodical triangular modulation signal (at about 1 Hz). This modulation is applied to another PZT stage (with Newport NPM140) on M3 in Fig. 2 in the main text. This mode is used for (i) the alignment for the parametric oscillation and (ii) the characterization of the longitudinal modes and oscillation threshold. Next, we move to the \textit{continuous operation}, where the OPO keeps the maximum power in one longitudinal mode with the servo controller. We define the oscillation threshold of the OPO as the pump power where all the oscillation peaks get invisible.

\subsection{Cavity modes in scanning operation}
A well-aligned system gives some resonant longitudinal down-converted modes in the scanning operation as shown in Fig. S\ref{fig:OPOScan}. Here, the cavity has two 90:10 output couplers, and the average pump power is 300 mW. The normalized signal power at the slow detector has separate peak structures with their intervals about a half wavelength ($\sim 0.79$ $\mu$m). We call them \textit{oscillation peaks} afterwards, and put numbers on the three highest peaks. The reference of the cavity length is set as the one at the top of the highest peak. The number of resonant peaks increases as the pumping power rises. Peak 1 is the degenerate peak and has the narrowest range of resonant cavity lengths in the three. It is probably because the degenerate peak has a narrower spectral range of the oscillating modes than a non-degenerate peak which has separate signal and idler spectral peaks. Peak 3 is a totally non-degenerate peak. We can see a small dip in peak 2, indicating that it has both degenerate and non-degenerate components \cite{paper:VMHV12}.

\subsection{Output power}
Fig. S\ref{fig:OPOPPOPW} presents the average output powers of the three strongest peaks in the continuous operation, as functions of the pump power. The signal power is measured with the output port from an output coupler. The mode of Peak 2 oscillates with a smaller pump power than peak 1 probably because peak 2 has a broader wavelength range in the optical spectrum. However, the mode of peak 1 has a larger nonlinear gain hence a large external quantum efficiency than that of peak 2. On the other hand, those for peak 2 and 3 are close. Considering the phase matching, this indicates that peak 1 is degenerate while the major components of peak 2 and 3 are non-degenerate. The efficiencies are 1.8, 1.3 and 1.2 \% for peak 1, 2 and 3, respectively. Such small values are because of a small reflection rate of 10 \% of the output port and the existence of two OCs in the system. In the case of only one output coupler, the threshold is 85 mW. Another factor in the restriction of an efficient conversion is the sum frequency generation, which consumes the pump and signal then generates a bright green beam in this experiment.

\subsection{Spectrum}
\begin{figure}[htbp]
\begin{center}
\includegraphics[width=10cm,trim = 0 20 0 0,clip]{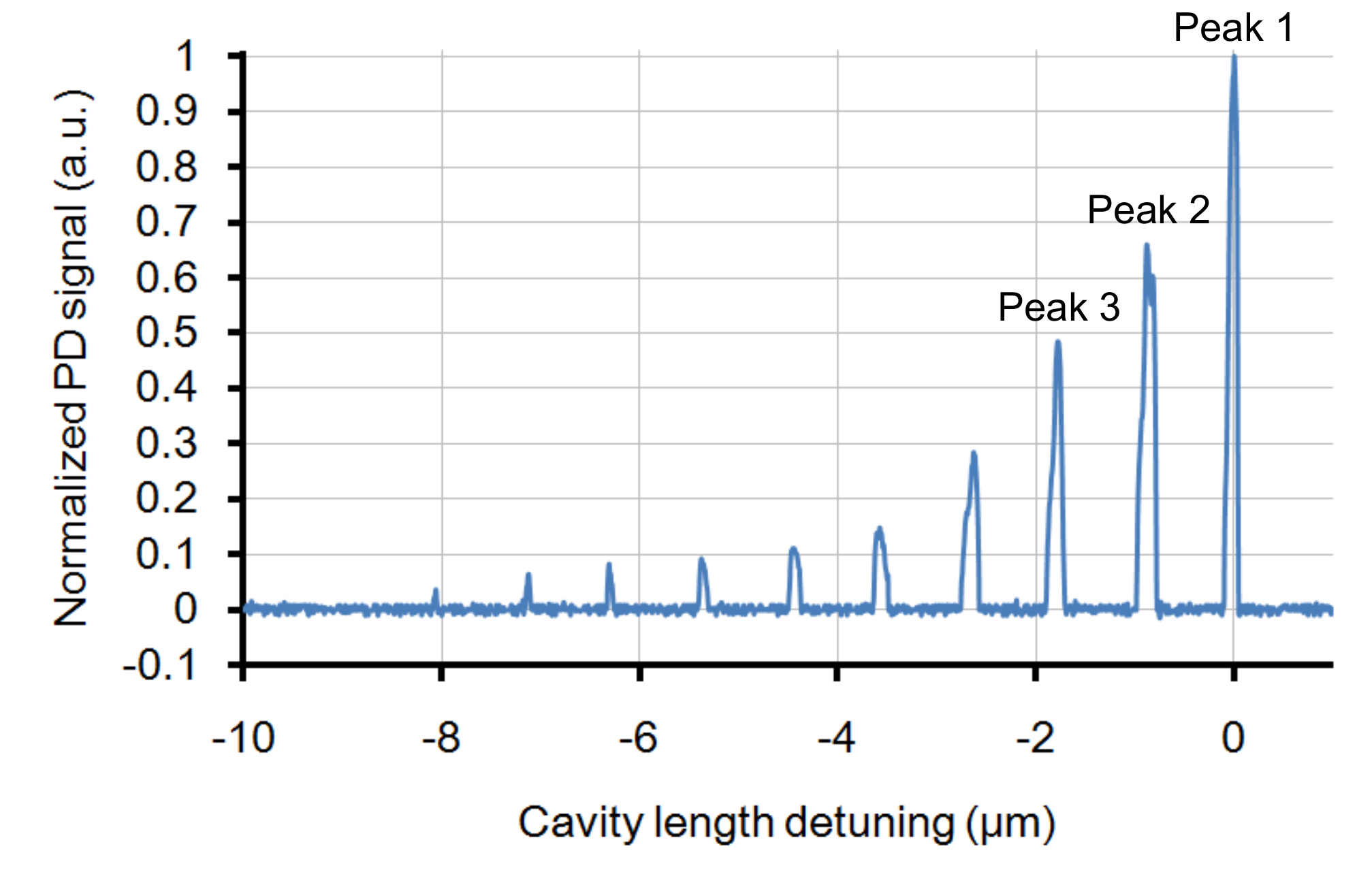} 
\end{center}
\caption{Average power of the OPO output dependent on the cavity length, showing longitudinal oscillation peaks. The average pump power is 300 mW. }\label{fig:OPOScan}
\end{figure}

\begin{figure}[htbp]
\begin{center}
\includegraphics[width=10cm]{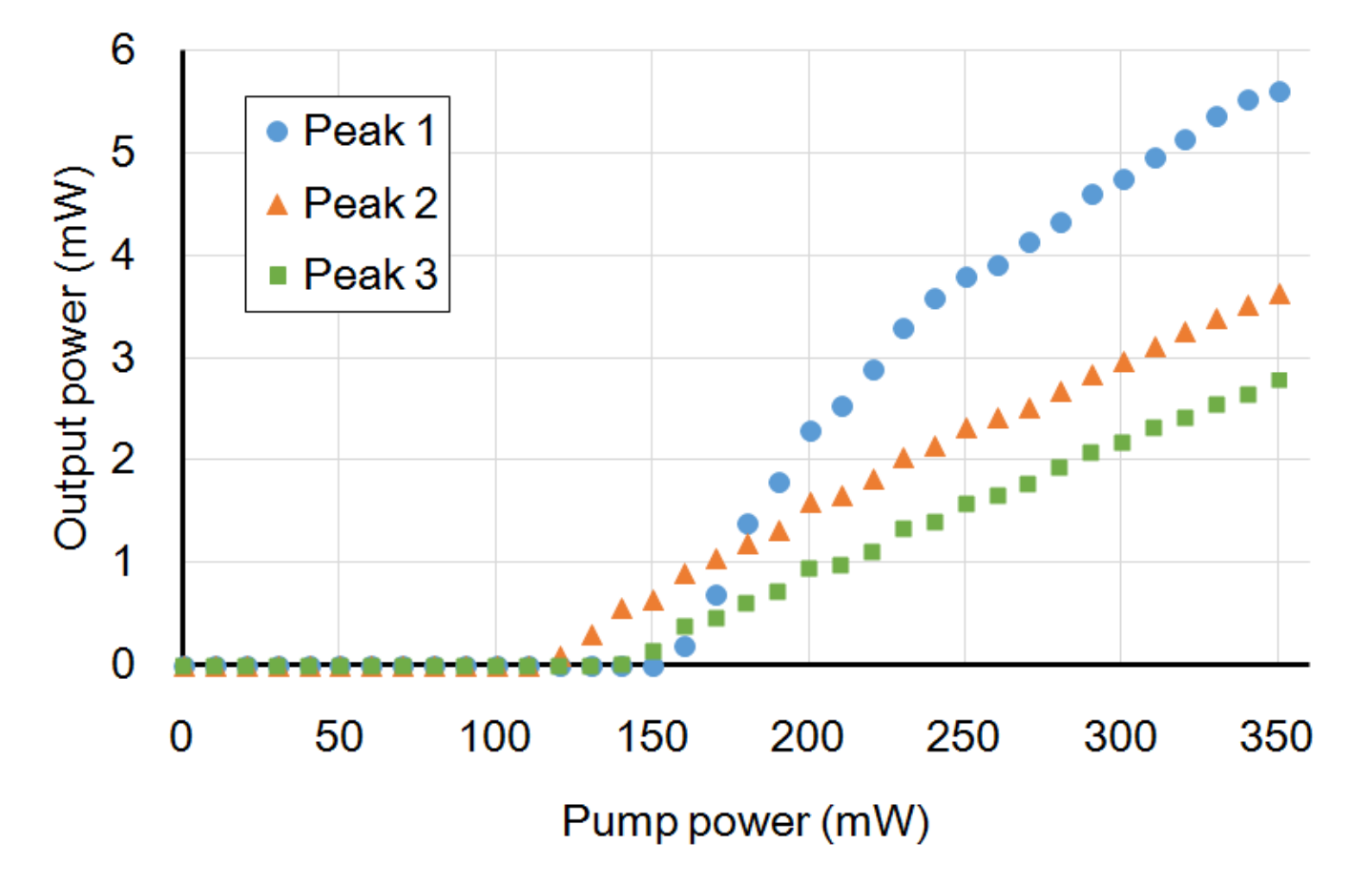}
\end{center}
\caption{Average output powers of the tops of the three oscillation peaks dependendent on the pump power. One of the two output couplers with 10\% reflection is used to take the signal.}\label{fig:OPOPPOPW}
\end{figure}

\begin{figure}[htbp]
\begin{center}
\includegraphics[width=10cm]{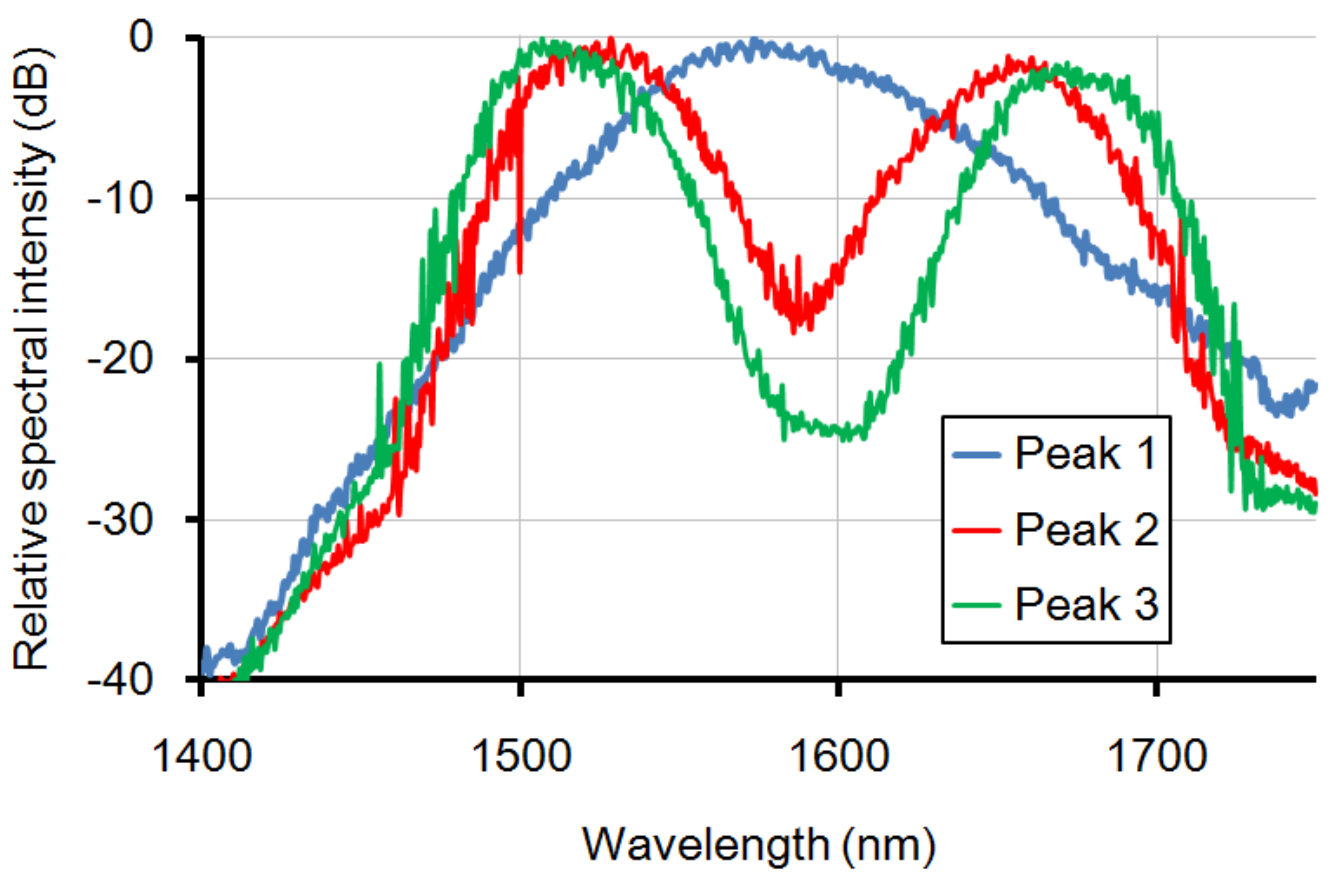}
\end{center}
\caption{Optical power spectra for the three oscillation peaks.}\label{fig:OPOSPPKS}
\end{figure}

The output beam is coupled to a multi-mode fiber and directed into an optical spectrum analyzer. The optical spectrum is measured as the average of 1000 sweep data and shown in Fig. S\ref{fig:OPOSPPKS} for the three oscillation peaks. We see that the spectrum for peak 1 is degenerate. Its peak wavelength and FWHM are 1574 nm and about 80 nm. Those for peak 2 and 3 have two separate spectral peaks, meaning that they are non-degenerate. Here, peak 2 has more spectral components including small spikes around the sub-harmonic wavelengths centered at 1588 nm. While the data is taken under the stable locking at the top of each oscillation peak, a large fluctuation sometimes makes the spectrum of peak 2 instantly switch to a single-peak structure like that for peak 1. This suggests that the second highest top in the right of peak 2 in Fig. S\ref{fig:OPOScan} is a degenerate or intermediate mode as the previous study referred to \cite{paper:VMHV12}.

\subsection{Pulse duration}
\begin{figure}[htbp]%
\begin{center}
\includegraphics[width=7cm]{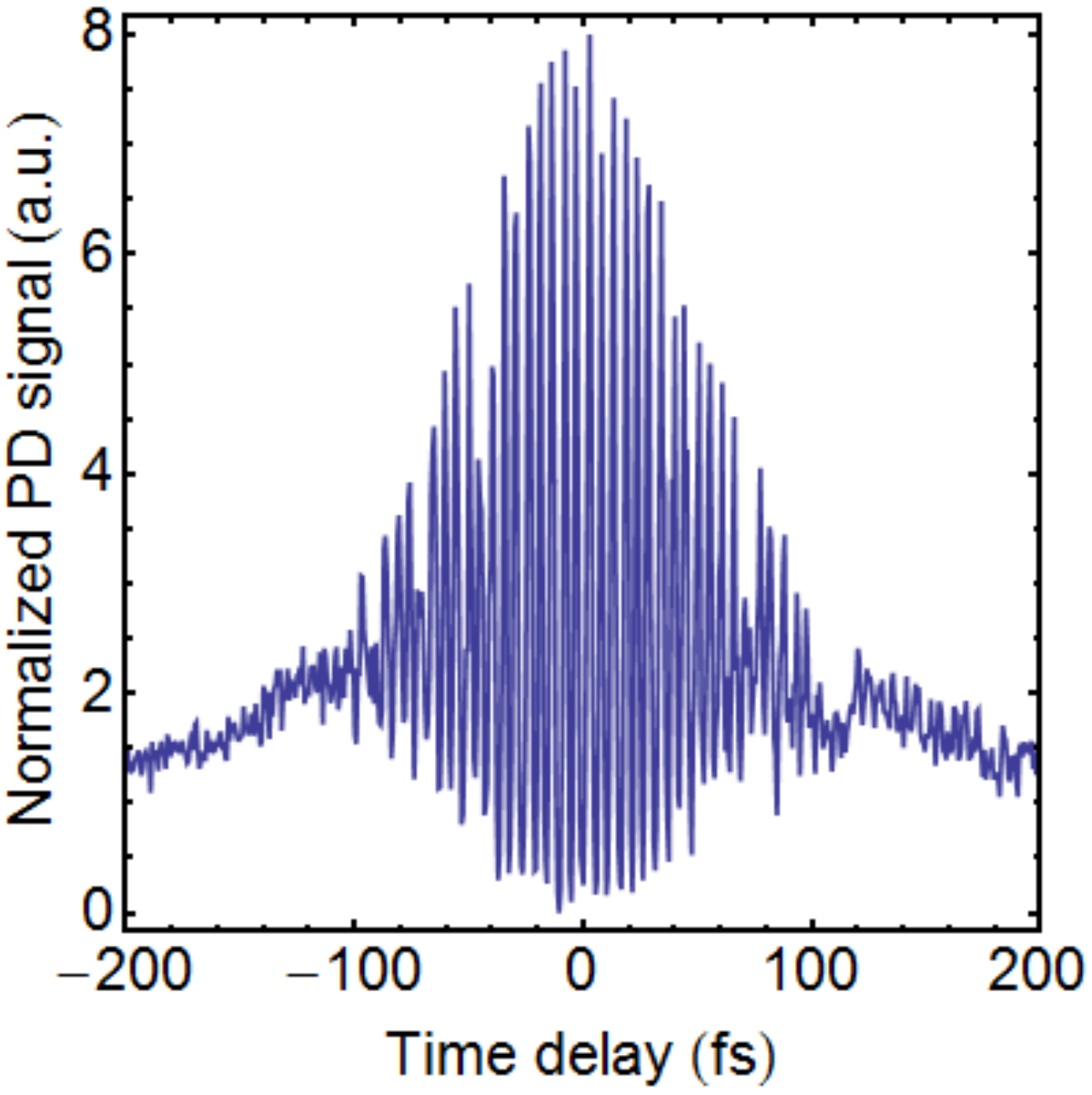}
\end{center}
\caption{Autocorrelation trace of the two-photon absorption for the degenerate OPO pulses.}\label{fig:OPOATCPK1}
\end{figure}

The temporal pulse duration for the degenerate signal mode (peak 1) is obtained by the autocorrelation measurement with a Michelson interferometer. Here, an interference fringe via the two-photon absorption in a Si detector (Thorlabs PDA36A) is depicted in Fig. S\ref{fig:OPOATCPK1}. The measured FWHM of the trace is 127 fs. When assumed as a sech$^2$ pulse, the sub-harmonic pulse has a duration of 80 fs \cite{book:Weiner09}, corresponding to about 15 optical cycles.

The spectral width of degenerate pulses in frequency is 9.6 THz, thus a pulse length of 32.7 fs is expected for a transform-limited sech$^2$ pulse \cite{book:Weiner09}. We presume that the deviation of the measured duration from that in the transform limit mainly comes from the GVD and chirping on the pump pulses at the power controlling part (HWP: Thorlabs AHWP10M-980, PBS: Newport UPBS-1). The duration of the pump pulses coming from OC2 has been measured with a GaAsP photodiode (Hamamatsu G6262) and found to be 44 fs, which is much larger than the value in the specification (15 fs). Another factor is the optical filter to pick only the signal pulses for the measurement.

\subsection{RF spectrum}
\begin{figure}[htbp]
\begin{center}
\includegraphics[width=12cm]{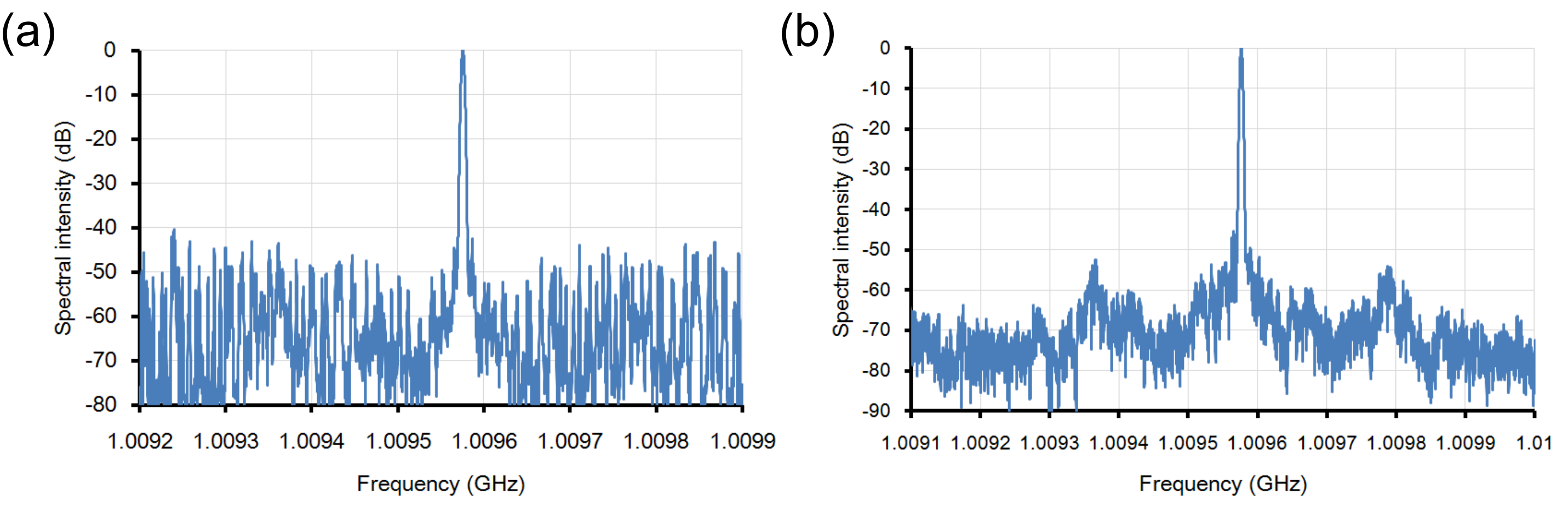}
\end{center}
\caption{RF spectra for the signal modes of the DOPO without delay lines. The peak frequency is 1.00956 GHz and corresponds to the pulse repetition frequency. The pump power is 300 mW. (a) The strongest degenerate mode (peak 1). (b) The second strongest non-degenerate mode (peak 2).}\label{fig:APP2DOPORF}
\end{figure}

Fig. S\ref{fig:APP2DOPORF} (a) and (b) show the RF spectra of the degenerate and non-degenerate oscillation peaks (peak 1 and 2) of the pulsed DOPO without delay lines. The spectral peak frequency reflecting the pulse repetition rate is 1.00957 GHz and is fluctuated by the order of 1 kHz. Peak 1 (Fig. S\ref{fig:APP2DOPORF} (a)) has larger background spectral components than peak 2 (Fig. S\ref{fig:APP2DOPORF} (b)). It is probably because the locking of peak 1 is more unstable due to a narrower oscillation peak in terms of the cavity detuning. Fig. S\ref{fig:APP2DOPORF} (b) (peak 2) has clear side peaks about $\pm 200$ kHz away from the main peak. This indicates the unstable operation of the oscillator because of the co-existence of a degenerate and non-degenerate modes, as was reported previously \cite{paper:Alireza12}.

\section{Complementary exprimental results of the 16-bit coherent Ising machine}
\subsection{Beam spot}
\begin{figure}[htbp]
\begin{center}
\includegraphics[width=10cm]{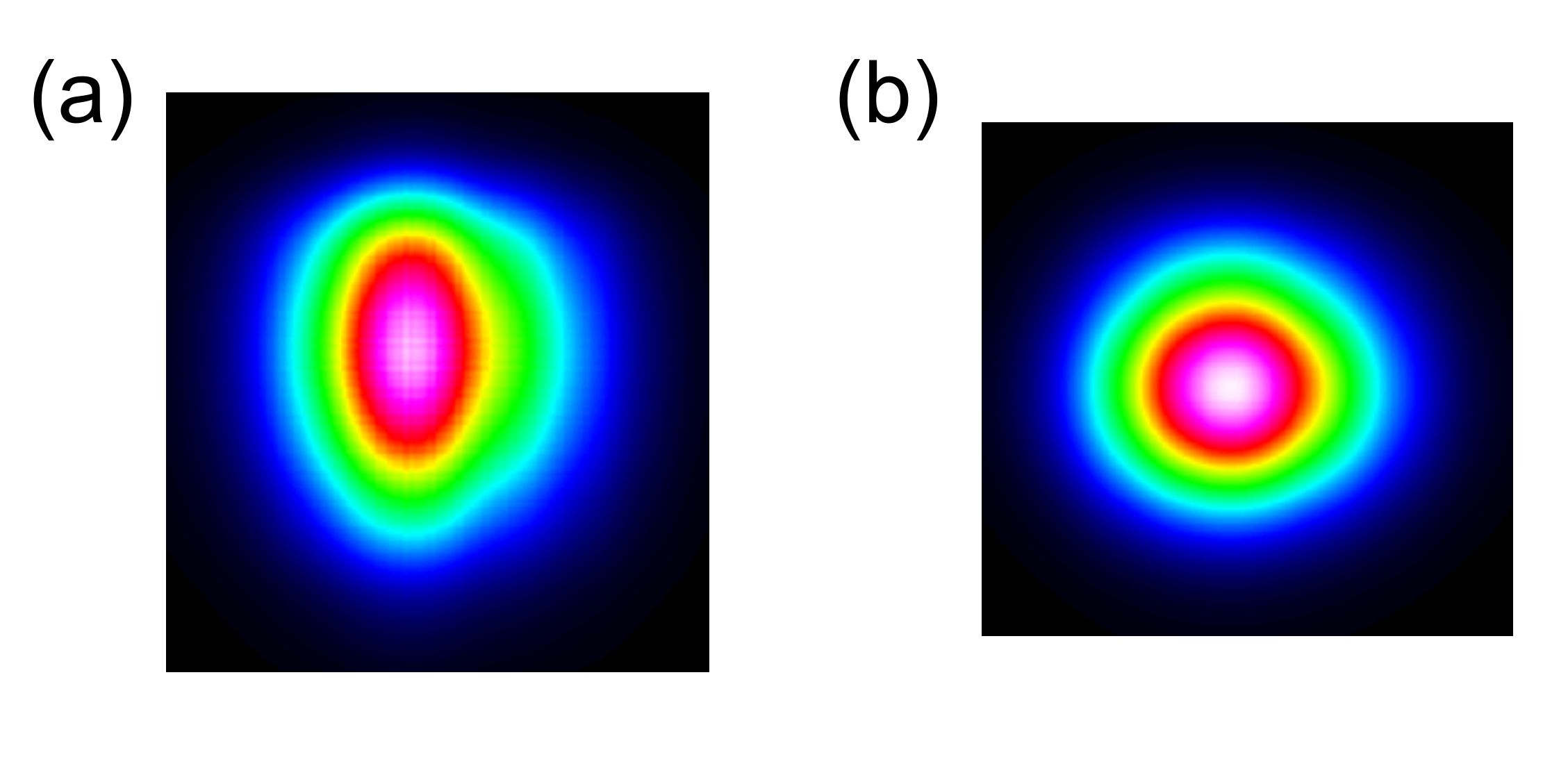}
\end{center}
\caption{Beam spots measured with a slit beam profiler for the DOPO with three delay lines. The system is strongly pumped and has degenerate modes in an odd and even frequency state. (a) Spot of an even frequency state (suitable for coherent computing). (b) One in an odd frequency state.}\label{fig:APP2DOPOSPOT}
\end{figure}

We show additional results of the experiment on the 16-pulse OPO Ising machine. Fig. S\ref{fig:APP2DOPOSPOT} displays the beam spots of the degenerate signal modes in the system with eight beamsplitters. A slit beam profiler is used for the measurement. The pump power is 900 mW, and the delay lines are blocked expect for an output port for the measurement. When many beamsplitters are introduced (for the couplings) and a high pumping power is used, some degenerate signal modes come to be available. Here, a degenerate mode is called an even frequency state if it keeps the fixed carrier phase with every round trip. On the other hand, the state suffering from a $\pi$ phase shift after a round trip is called an odd frequency state. Fig. S\ref{fig:APP2DOPOSPOT} (a) is the spot for an even frequency state, and (b) is one for an odd frequency state. The spot size (radius) in the horizontal ($w_x$) and vertical ($w_y$) directions are $(w_x,w_y) =$ (1.7 mm, 1.9 mm) and (1.7 mm, 1.5 mm) for Fig. S\ref{fig:APP2DOPOSPOT} (a) and (b), respectively. The shape of the spot for the odd frequency state is closer to that of the pump beam indicating that the odd frequency state is closer to the stablest condition for the main OPO ring cavity. Note that repetitive overlaps of signal and pump pulses are important for the resonance and affected by the effective main cavity length varying with the number of input and output couplers.


\subsection{Random phase states of each pulse}
\begin{figure}[htbp]%
\begin{center}
\includegraphics[width=12cm]{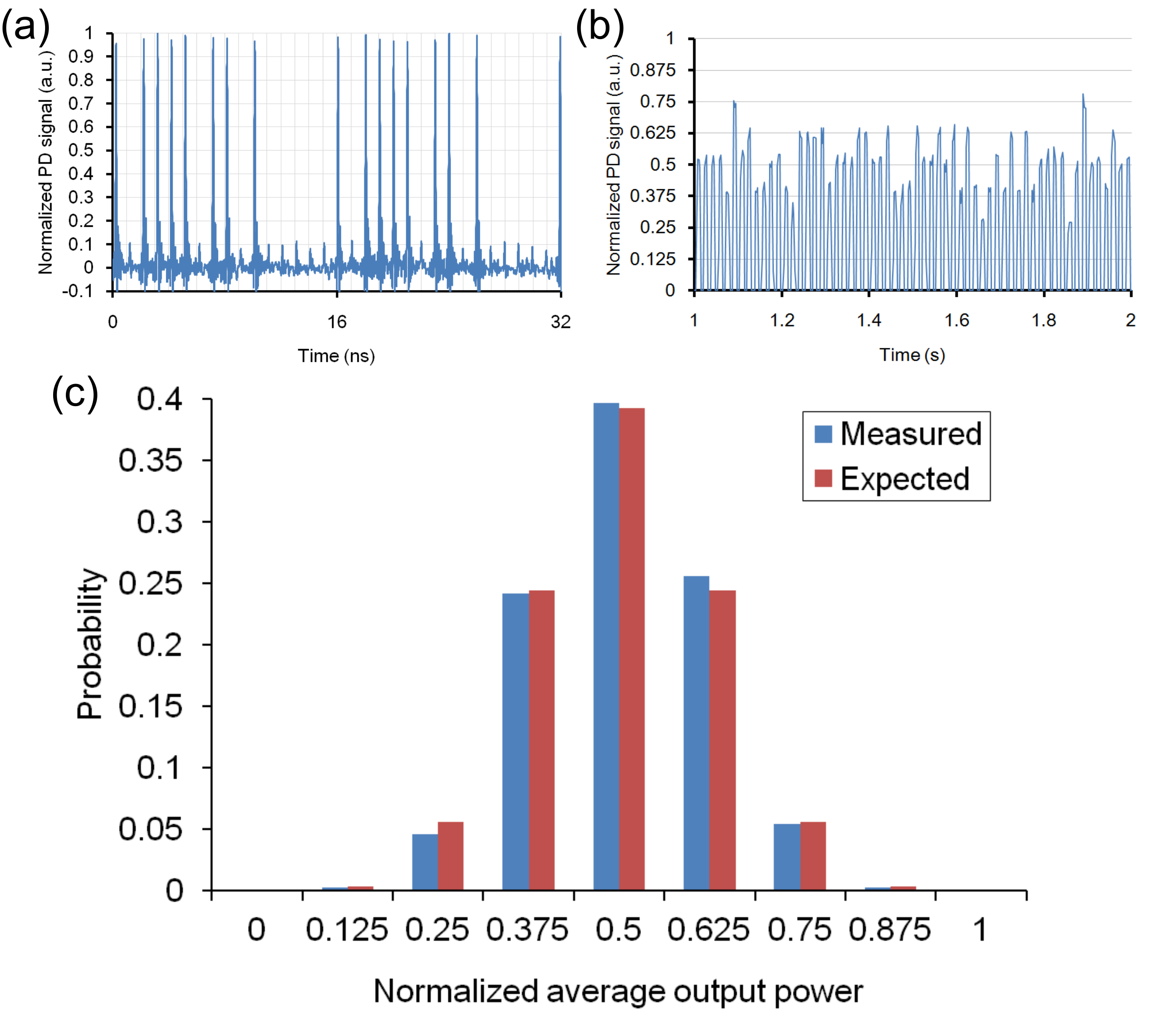}
\end{center}
\caption{Detector readouts for the system without the mutual injection. (a) An example of the interferometer pulse patterns measured with the fast detector. (b) The average output power under the chopping with a frequency of 60 Hz. (c) Distribution of the average output power out of 1500 trials.}\label{fig:16OPORANDOM}
\end{figure}

For computation on the Ising model, we drive the system with the DOPO pulses in an even frequency state. Here, we introduce the system with six output couplers in the main cavity, and all the delay lines are blocked. The pump power is 2.5 times the oscillation threshold. Fig. S\ref{fig:16OPORANDOM} (a) displays an example of the interferometer output pulse patterns measured with the fast detector. The period of 16 pulses ($\sim 16$ ns) in the pattern means the system contains as many pulses. The binary intensity levels confirm that the signal pulses are degenerate. An even number of high-intensity pulses (eight here) indicates that the DOPO pulses have an even frequency state. The power fluctuation of about 5 \% mainly reflects that of the pump laser and the fine structure of the cavity length dependence of the DOPO pulse intensity. We have observed various pulse patterns in an even frequency state via automated measurements.
When we consider to test the randomness of the binary phase states, the possible output pulse patterns is too many to list. Here, we use the average power of the interferometer output under the chopping with a frequency of 60 Hz. Fig. S\ref{fig:16OPORANDOM} (b) is the average power normalized by the twice of the mean of the DOPO signals (with the chopper open). Here, the number of high-intensity pulses are even thus the normalized average power is discretized to $n/8$ $(n = 0, 1, \ldots, 8)$. The change in the average power at restarts is seen, indicating that the DOPO pulses are damped at each blocking of the pump by the chopper.
Fig. S\ref{fig:16OPORANDOM} (c) presents the probabilities of detecting the normalized output power levels out of 1500 independent trials. The measured data is close to that in the case of random phase states calculated by the brute-force counting. The observation here supports the randomness and independence of the binary phase states of each DOPO pulse.

\subsection{Phase locking by injection with a single delay line}
\begin{figure}[htbp]
\begin{center}
\includegraphics[width=12cm]{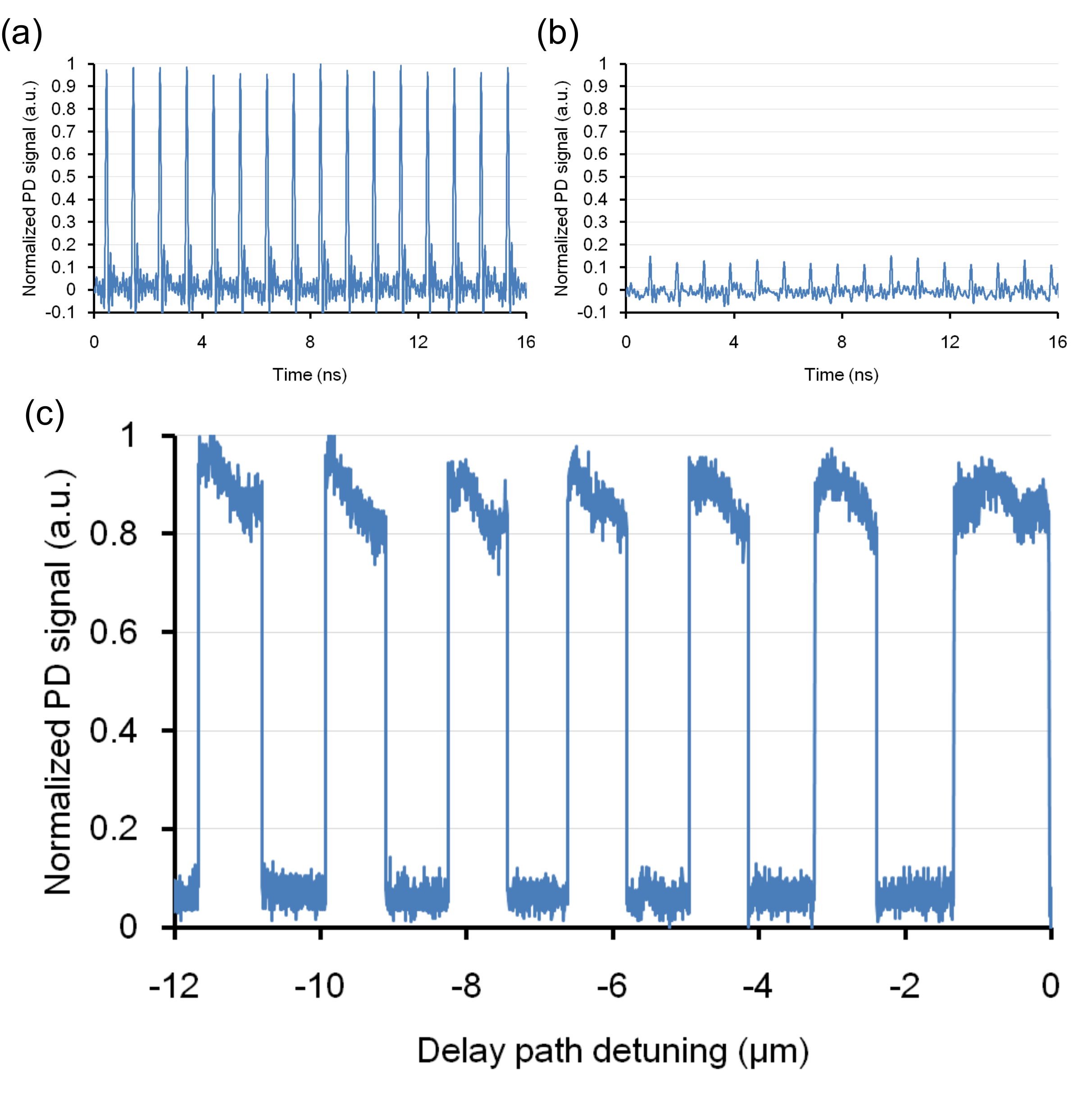}
\end{center}
\caption{The temporal interferometer output power for the system with a single delay line. (a) Fast detector readout in the case of in-phase unidirectional couplings. (b) That for out-of-phase unidirectional couplings. (c) Slow detector readout under the scanning of the delay path length.}\label{fig:16OPO1DL}
\end{figure}

Here, we introduce the shortest delay line and run the system without the chopping. Fig. S\ref{fig:16OPO1DL} (a) and (b) depict the fast detector outputs of the signal interference in this case. Tuning and locking the length of the delay line with the pump interference enable us to obtain the DOPO pulse with the in-phase and out-of-phase order at steady states as represented in Fig. S\ref{fig:16OPO1DL} (a) and (b), respectively. The locking points for the in-phase and out-of-phase order appear alternately because the center wavelength of the pump is about twice longer than that of the sub-harmonic mode. Fig. S\ref{fig:16OPO1DL} (c) is a slow detector readout of the signal interferometer output when the delay path length is scanned with a PZT. The high- and low-level signal correspond to the states in the in-phase and out-of-phase order (Fig. S\ref{fig:16OPO1DL} (a) and (b)). The maximum path detuning computed by the specification of the PZT is 12 $\mu$m. It corresponds to about 7.6 cycles for the carrier wave at the sub-harmonic wavelength 1.574 $\mu$m, giving a good agreement with the number of interference cycles in the figure. The slopes in the graph are possibly because of different extents of overlaps between the pulses for different locking points, in terms of the peak amplitudes and chirping. The graph has been obtained by decreasing the delay path length. When acquired by increasing the length, the signal can give the levels corresponding to the erroneous states.

\subsection{Erroneous states}
\begin{figure}[htbp]
\begin{center}
\includegraphics[width=12cm]{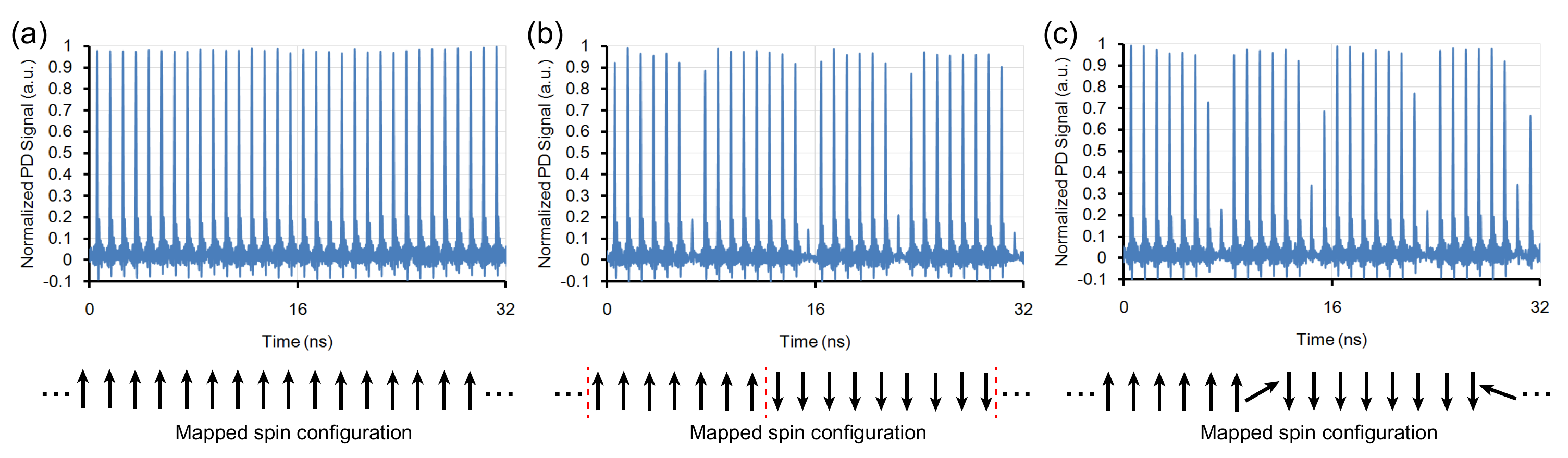}
\end{center}
\caption{Fast detector readouts of the interferometer output for the 16-DOPO pulse system with the bidirectional in-phase couplings in the form of the one-dimensional ring. The performance depends on the temporal inteference between the cavity pulses and injected pulses. (a) The stablest state corresponding to the ground state of the ferromagnetic Hamiltonian. (b) An excited state with two domains of in-phase pulses. (c) An excited state seen when the locking points of the delay lines are not good. Finite residual coupling phases rotate the phases of the boundary pulses and make the output peak powers non-disrete around them. The mapped spin configuration for each output is added.}\label{fig:16OPO2DLF}
\end{figure}

\begin{figure}[htbp]
\begin{center}
\includegraphics[width=12cm]{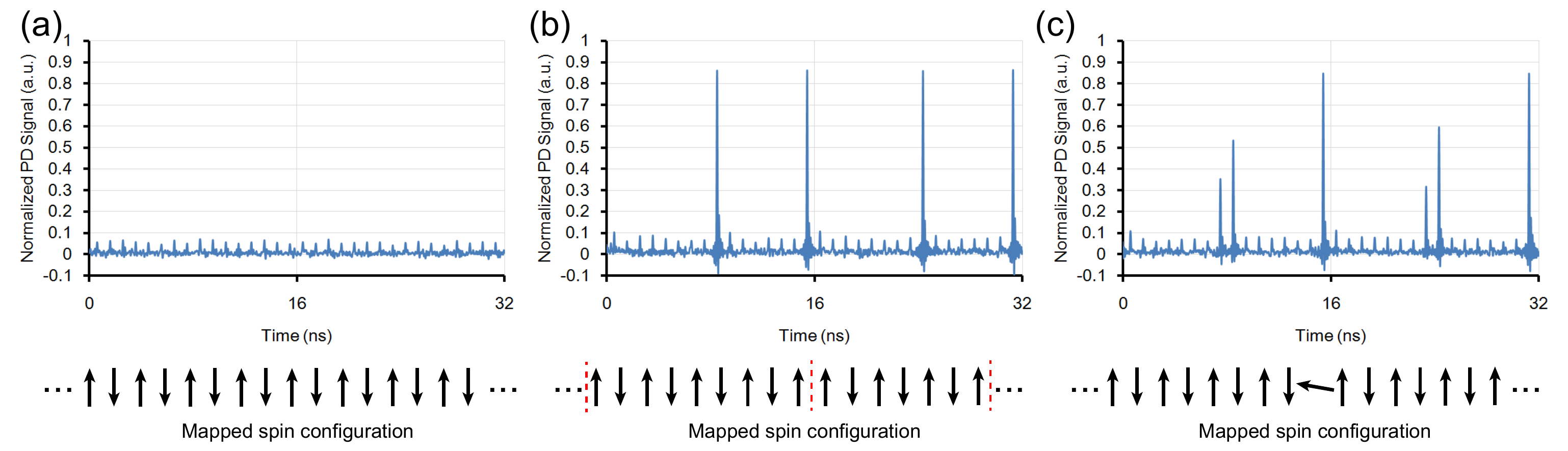}
\end{center}
\caption{Fast detector signal of the  output for the 16-DOPO system with the bidirectional out-of-phase couplings in the one dimensional ring. (a) The state with the minimum gain, corresponding to the ground state of the anti-ferromagnetic Hamiltonian. (b) An excited state with two clusters of out-of-phase pulses. (c) An excited state for the case of bad locking points of delay lines. The mapped spin configuration for each output is added.}\label{fig:16OPO2DLAF}
\end{figure}

Here, we show a few of examples of the erroneous output states under suboptimal conditions. Fig. S\ref{fig:16OPO2DLF} presents fast detector outputs from the interferometer in the case of in-phase mutual injections emulating the ferromagnetic Ising model ($J_{i \, i+1} \approx  J_{i+1 \, i} > 0$). We add an schematic mapped spin configuration expected from each pulse pattern. The phase states with the global in-phase order identified by Fig. S\ref{fig:16OPO2DLF} (a) correspond to the two ground states and most probable. However, some excited phase states can be detected with a finite probability when the lengths of the delay lines are not optimum. Major erroneous states have two domains of in-phase DOPO pulses corresponding to two ferromagnetic domains as shown in Fig. S\ref{fig:16OPO2DLF} (b) and (c). The real-time observation of the output indicates that the domains are relatively stable when the locking point is nearly optimum (Fig. S\ref{fig:16OPO2DLF} (b)). Meanwhile, when the locking point is off and the success probability gets worse, we see pairs of sequential output pulses whose powers are not binary (Fig. S\ref{fig:16OPO2DLF} (c)). Their powers also vary in time, suggesting that the phase of the pulse at the domain boundary gets unstable due to the residual coupling phases by the delay lines. Same applies also to the case of the out-of-phase mutual injections (emulating the anti-ferromagnetic Hamiltonian $J_{i \ i+1} \sim J_{i+1 \ i} < 0$) as shown in Fig. S\ref{fig:16OPO2DLAF}. The stablest phase-state configuration has the global out-of-phase order (Fig. S\ref{fig:16OPO2DLAF} (a)), while major excited phase states like in Fig. S\ref{fig:16OPO2DLAF} (b) indicate two anti-ferromagnetic domains. When the locking points are not good ones, the phases of the boundary pulses rotate and output peak powers around them get non-discrete (Fig. S\ref{fig:16OPO2DLAF} (c)).

\begin{figure}[htbp]
\begin{center}
\includegraphics[width=12cm]{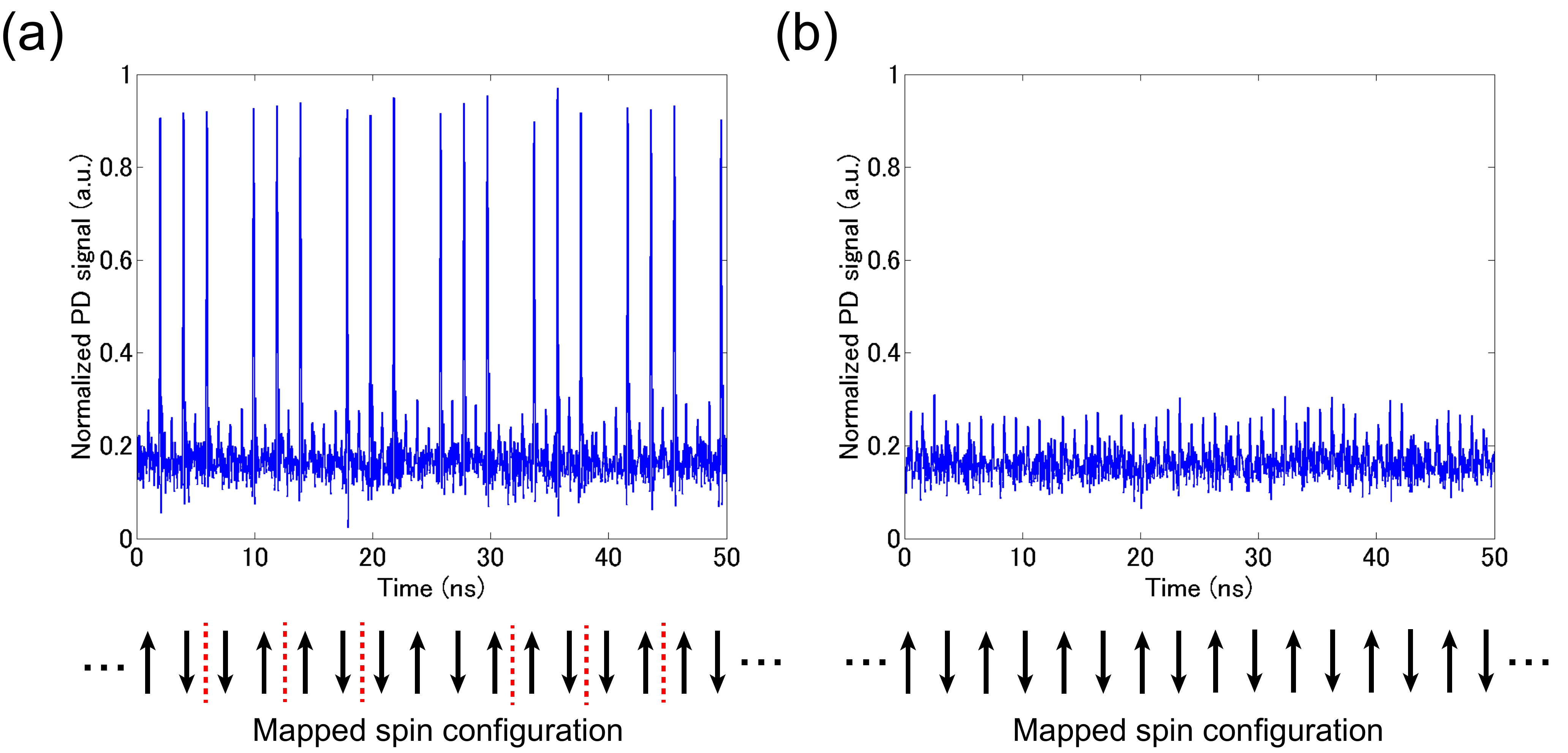}
\end{center}
\caption{The interferometer output pulse patterns for the two failed cases both of which correspond to local minima. (a) reflects the couplings along with the diameter chords of the graph. (b) is affected by the anti-ferromagnetic order of the 1-D ring structure. Red dashed lines for the mapped spin configuration in (a) mean the frustration between adjacent spins. Note that all the couplings along with the diameter chords ($J_{n \ n+8 (mod \, 16)}$) are frustrated in (b), although it is not directly reflected in the output signal showing the interference between adjacent pulses.}\label{fig:16OPOLocalMinima}
\end{figure}

Finally, we refer to the cubic graph problem. Fig. S\ref{fig:16OPOLocalMinima} depicts the interferometer outputs obtained as the two erroneous states with a nearly best locking point. These are the only failed cases under this condition, out of 1000 runs. Both are ones of the 34 local minima. Fig. S\ref{fig:16OPOLocalMinima} (a) reflects $J_{n \ n+8 (mod \, 16)}$ and the corresponding state has six frustrated couplings. Fig. S\ref{fig:16OPOLocalMinima} (b) indicates the anti-ferromagnetic order along with the one-dimensional ring and gives eight frustrated edges along with the diameter chords. These results of local minima mean that the erroneous states well reflect the interaction between artificial spins. Meanwhile, they are not expected in our simulation and hence show the sensitiveness of the perfomance to experimental defectiveness.

\newpage


\end{document}